\documentclass[aps,prstab,twocolumn,groupedaddress,
nofootinbib,floatfix]{revtex4-1}
\usepackage{graphicx}
\usepackage{dcolumn}
\usepackage{etoolbox, ifpdf}
\usepackage[usenames,dvipsnames]{color}
\usepackage{soul}
\usepackage{relsize}
\usepackage{booktabs}
\usepackage{amsmath}
\usepackage{hyperref}
\hypersetup{
    pdfstartview={FitH},
    colorlinks=false,
    pdfborder={0 0 0},
    breaklinks=true
}
\usepackage{breakurl}

\begin{document}

\title{Low-emittance tuning at the Cornell Electron Storage
Ring Test Accelerator}
\author{J. Shanks}
\email{js583@cornell.edu}
\author{D.L. Rubin}
\author{D. Sagan}
\affiliation{CLASSE, Cornell University, Ithaca, New York 14853, USA}
\date{\today}

\begin{abstract}
    In 2008 the Cornell Electron/Positron Storage Ring (CESR) was
    reconfigured from an electron/positron collider to serve as a testbed for
    the International Linear Collider (ILC) damping rings. One of the primary
    goals of the CESR Test Accelerator (CesrTA) project is to develop
    a fast low-emittance tuning method which scales well to large rings such
    as the ILC damping rings,  and routinely achieves a vertical emittance
    of order $10\,\textrm{pm}$ at 2.085~GeV.
    This paper discusses the tuning methods
    developed at CesrTA to achieve low-emittance conditions.
    One iteration of beam-based measurement and correction requires
    about 10 minutes.
    A minimum vertical emittance of $10.3$ (+3.2/-3.4)$^{\;sys}$
    ($\pm$0.2)$^{stat}$~pm has been achieved at 2.085~GeV.
    In various configurations and beam energies the correction technique
    routinely achieves vertical emittance around $10$~pm after correction,
    with RMS coupling $< 0.5\%$. The measured vertical dispersion is
    dominated by beam position monitor systematics.
    The propagation of uncertainties in the emittance
    measurement is described in detail.
     Simulations modeling the effects of
    magnet misalignments, BPM errors, and emittance correction
    algorithm suggest the residual vertical emittance measured at the
    conclusion of the tuning procedure is dominated by sources other
    than optics errors and misalignments.
\end{abstract}

\maketitle

%#######################################

\section{Introduction}

    In 2008 the Cornell Electron Storage Ring (CESR) was reconfigured from an
    electron/positron collider to the CESR Test Accelerator (CesrTA)
    \cite{PAC09:FR1RAI02, ICFABDNL50:11to33, PAC09:WE6PFP103},
    a testbed for the International Linear Collider (ILC) damping
    rings \cite{ilc:tdr}. Parameters for the CESR storage ring are shown in
    Table \ref{tab:cesr_params}. One of the primary objectives of the
    CesrTA program is to develop low-emittance tuning methods for
    the ILC damping rings.

    \begin{table}[htb]
        \caption{Parameters of the CESR electron/positron
        storage ring.}\label{tab:cesr_params}
        \begin{tabular}{lcc}
            \toprule[1pt]
            \addlinespace[2pt]
            \textbf{Parameter} & \textbf{Value} & \textbf{Units}\\
            \midrule[0.5pt]
            \addlinespace[2pt]
            Circumference & 768.4 & m \\
            Energy      & 2.085 (1.5-5.3)& GeV \\
            Lattice Type      & FODO & \\
            Tunes ($Q_x, Q_y$)& (14.59, 9.63) & \\
            Symmetry          & $\approx$ Mirror & \\
            H / V Steerings & 55/58 & \\
            Quadrupoles & 105 & \\
            Skew Quadrupoles & 27 & \\
            Damping Wigglers & 12 & \\
            Wiggler $B_{max}$ & 1.9 & T \\
            Position Monitors & 100 & \\
            ${\epsilon_x}^{geometric}$ & 2.7 & nm \\
            ${\epsilon_y}^{geometric}$ (target) & 10 & pm \\
            \bottomrule[1pt]
        \end{tabular}
    \end{table}

    By far the most common
    tool for linear optics correction is Orbit Response Matrix (ORM)
    analysis, specifically Linear Optics from Closed
    Orbits (LOCO) \cite{NIMA:LOCO, SLACPUB:9464}. In particular, LOCO
    has been used as the cornerstone for corrections at both the
    Swiss Light Source (SLS) and the Australian Synchrotron, where
    vertical emittances of order 1~pm have been reported
    \cite{NIMA:SLS, PRSTAB14:012804}.

    However, the time required for measuring the response matrix scales
    linearly with the number of correctors. The Australian Synchrotron
    has demonstrated an acquisition rate of order 10 seconds per
    corrector. Assuming the ILC damping rings will be capable of the
    same acquisition rate, simply measuring the response matrix for all 800
    steerings would take several hours, and thus response matrix analysis
    is deemed prohibitively slow for the ILC damping rings.

    The tuning algorithm developed for CesrTA was required to be
    fast, and scale well to large rings such as the ILC damping rings.
    The correction procedure takes less than 5 minutes to acquire
    a full data set, where the duration time is limited by the slew rate of
    the superconducting RF cavities for dispersion measurements. One
    correction iteration  (measure, compute corrections, load corrections, and
    remeasure) takes around 10 minutes.
    Data acquisition is fully parallelized, with pre-processing done
    on beam position monitor (BPM) modules, one per BPM.
    Measurement time for the CesrTA algorithm scales
    independently of number of BPMs, and does not depend on the number of
    correctors.

    The algorithm may also prove useful to other storage
    rings. Betatron phase measurements are significantly faster than
    traditional response matrix analysis, allowing for less time to be
    spent on optics correction. The measurements may be performed using a
    witness bunch, exciting and measuring only a single bunch in a
    fully-loaded machine. Additionally, measurements such as betatron phase
    and coupling which resonant excitation do not require changing the
    machine conditions, minimizing hysteresis.

    This paper describes the optics correction procedure developed
    at CesrTA that meets these requirements. Experimental results,
    with detailed propagation of uncertainties, are presented.
    Also discussed are simulations of the correction procedure, which have
    been essential to understanding measurement
    systematics and recognizing that the residual vertical emittance
    is dominated by sources other than optics errors and misalignments.

%---------------------------------------------------
\section{Motivation for Beam-Based Emittance Tuning}

    For the ILC, the quantity of interest is not the projected vertical
    beam size but rather the emittance $\epsilon_b$ of the
    vertical-like normal mode, called the ``$b$-mode.'' In principle, the beam
    could be intentionally coupled in the damping rings in order to reduce
    collective effects, and decoupled in the extraction line, so long as the
    $b$-mode emittance is preserved. The decomposition into normal
    modes has been discussed elsewhere \cite{PhysRevSTAB.9.024001,
    PhysRevLett.111.104801, PhysRevSTAB.15.124001},
    and therefore will not be covered here.

    The primary static
    contributions to $\epsilon_b$ in a planar ring are tilted and
    vertically-offset quadrupoles, and rolled dipoles. Tilted quadrupoles
    couple horizontal and vertical motion which couples photon emission
    in the horizontal plane to the b-mode. Vertical quadrupole offsets and
    dipole rolls introduce vertical kicks, generating vertical dispersion and
    thus vertical emittance. Additional sources of vertical emittance
    include time-dependent variations associated with
    line voltage, ground motion, and feedback systems, which contribute
    kicks to the beam in various ways, and current-dependent effects such
    as intra-beam scattering (IBS).

    Without beam-based corrections of dispersion and coupling, the vertical
    emittance would be limited by the quality of survey and
    alignment.

    Survey and alignment is accomplished primarily using a Leica AT402
    Absolute Laser Tracker for establishing the reference network and a
    Leica DNA03 digital level for elevation runs. Establishing the
    reference network is done via "free stationing," with over 100
    stations in the network. The reference network consists of a triplet
    of reference targets--attached to the inner tunnel wall,
    outer tunnel wall, and
    embedded in the floor--approximately every 8 to 10 meters in the
    tunnel arcs, and less regular, but on average equally dense wall,
    floor, and ceiling targets in the long-straight sections, flares,
    and main south areas. Typical combined (bundled) 1$\sigma$ uncertainties
    for the reference targets are on order 35~$\mu$m for height (z).
    Magnets are surveyed into place to better than 100~$\mu$m of measured
    versus ideal positions using fiducials mechanically aligned to their
    irons.  The reference network and magnets are resurveyed regularly.
    Comparisons of reference target variations from survey to survey
    establish the uncertainties used in the analysis.

\begin{figure}[tbh]
        \includegraphics[width=3.25 in]{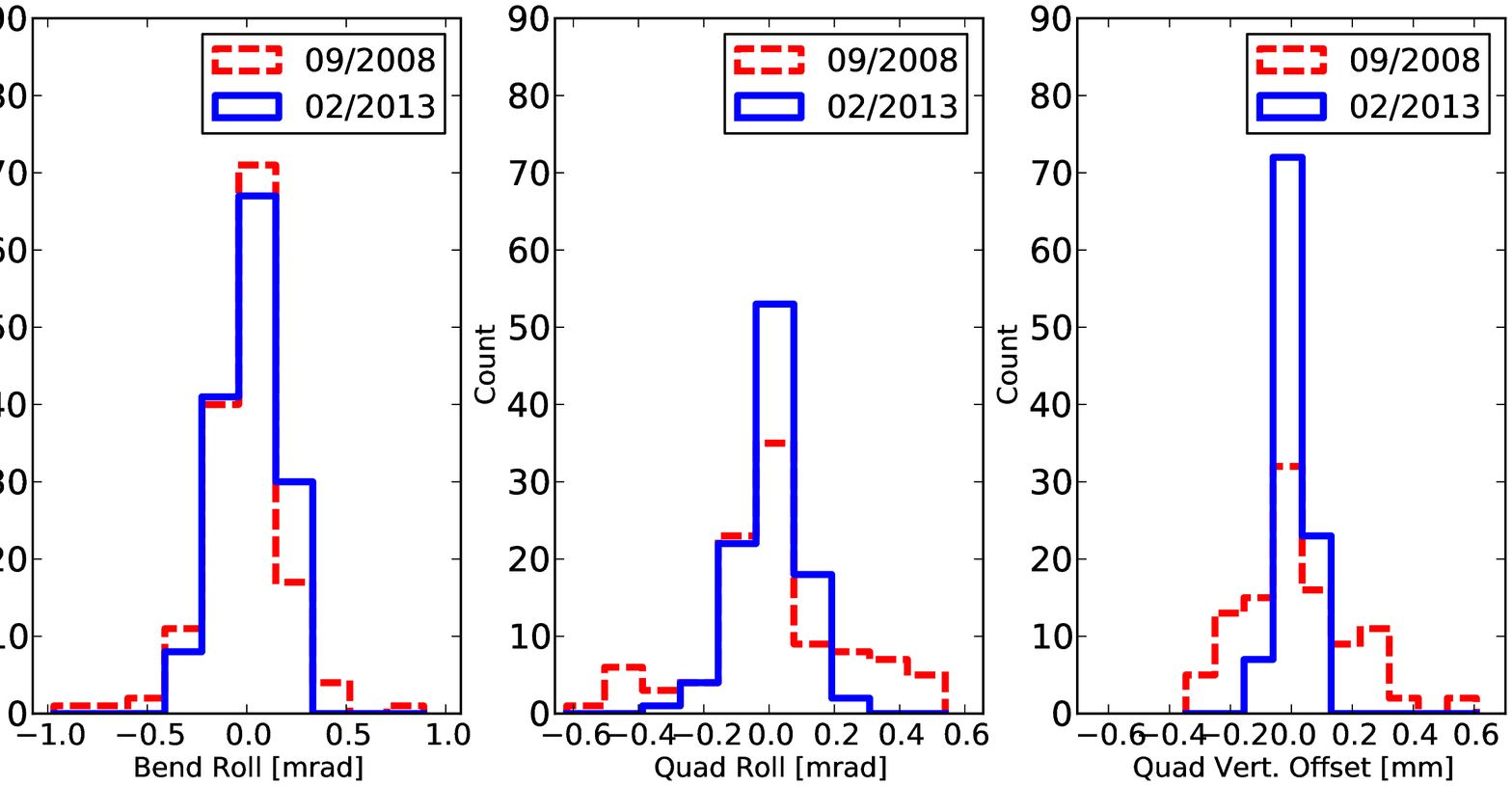}
        \caption{Survey and alignment results for CesrTA as of December
        2012 CesrTA run, compared to alignment in September 2008
        at the start of the CesrTA program. Left to right: dipole roll,
        quadrupole tilt, and quadrupole vertical offset.}
        \label{fig:survey_alignment}
    \end{figure}

    The measured distributions of surveyed
    quadrupole and dipole offsets and tilts for CESR are shown in Fig.
    \ref{fig:survey_alignment}. The root mean square (RMS) of the position
    and tilt measurements are summarized in Table
    \ref{tab:cesrta_sim_misalignments},
    and include an estimated $100\,\mu \textrm{m}$ uncertainty in the
    displacement of the magnetic center from physical center of the magnet.

\begin{figure}[tbh]
        \includegraphics[width=3 in]{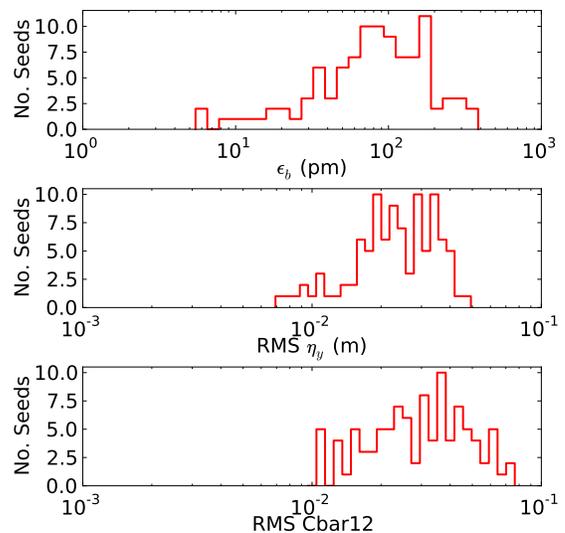}
        \caption{Resulting distributions of vertical emittance, vertical
        dispersion, and coupling when applying random distributions of
        errors at the amplitudes specified in Table
        \ref{tab:cesrta_sim_misalignments},
        along with systematic and random multipoles specified in Table
        \ref{tab:cesrta_sim_multipoles}. }
        \label{fig:no_correction}
    \end{figure}

    Simulations using random distributions of magnet, beam position monitor,
    and multipole errors consistent with measurements
    (summarized in the Appendix) have
    been used to study the effect of these errors on the vertical emittance.
    Repeating for 100 random sets of magnet errors,
    the resulting distributions of emittance, dispersion, and coupling yield
    statistical information about the probability of achieving the target
    emittance, and are shown in Fig. \ref{fig:no_correction}. The coupling
    is characterized using the $\bar{C}$ coupling matrix, an extension of the
    Edwards and Teng formalism \cite{IEEETransNuclSci:EdwardsTeng} and defined
    in \cite{PRSTAB2:074001}. In particular, only the out-of-phase
    coupling matrix element $\bar{C}_{12}$ is considered, for reasons discussed
    in Section \ref{sec:meas_techniques}.

    % /home/shanksj/CesrTA/ring_ma2_jobs/20130701_thesis_omni/baseline/tracking/*

    Without any beam-based corrections, simulations show that out of
    100 random seeds, only three yielded the target vertical
    emittance of 10~pm; the mean vertical emittance of the 100 seeds is
    $104~\textrm{pm}$. It is evident that the survey and alignment
    techniques used are insufficient by themselves to reach the CesrTA
    emittance target. Some form of beam-based correction is clearly
    required in order to achieve and maintain low-emittance operating
    conditions.

\section{Measurement Techniques}\label{sec:meas_techniques}

\begin{figure}[tbh]
        \includegraphics[width=3 in]{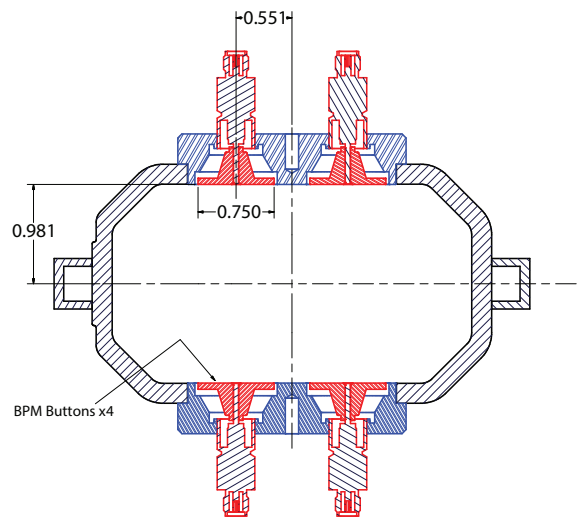}
        \caption{Cross-section of a CESR BPM. Dimensions are in inches.}
        \label{fig:cesr_bpm}
    \end{figure}

    Beam position monitors (BPMs) are used to collect data for most
    beam-based optics characterization techniques used in emittance
    tuning at CesrTA. CESR is instrumented with 100 button-style peak-detection
    BPMs. A cross-section of a typical CESR BPM is shown in Fig.
    \ref{fig:cesr_bpm}. New electronics, developed in-house, were installed
    in 2009 \cite{IPAC10:MOPE089}. The new BPM system is capable of
    bunch-by-bunch, turn-by-turn readout for bunch spacings $\ge$4ns with
    a buffer of 300,000 bunch-turns. At each BPM, all four button channels are
    read out by separate controller cards, therefore channel-to-channel
    crosstalk is minimized. Bunch-to-bunch cross-talk is below 4\% after
    4ns, and is effectively zero after 50ns; there is no turn-to-turn
    cross-talk. Single-turn orbit reproducibility is measured to be
    $7~\mu\textrm{m}$ for consecutive turns;
    as this is determined from beam-based measurements,
    it includes not only the sensitivity of the BPM, but also all contributions
    such as electronic stability and beam pipe vibration.
    Depending on the user's request for data, some level of
    pre-processing is done onboard the BPM electronics before committing data
    to file, or the raw bunch-by-bunch turn-by-turn button signals are written
    directly to file.

    BPMs are used to measure closed orbit,
    betatron phase and coupling, and dispersion. Turn-by-turn trajectory data
    is also used for BPM calibrations. For all beam-based measurements
    in low-emittance tuning at CesrTA, a single bunch of 0.8~mA =
    $1.3\times10^{10}$ particles is used. Therefore, bunch-to-bunch effects
    do not contribute to emittance measurements.

    The closed orbit at each BPM is measured by averaging 1024 turns
    of turn-by-turn bunch trajectory data onboard the BPM modules. A
    closed orbit measurement takes roughly 5 seconds, with measurement
    reproducibility of around $2~\mu\textrm{m}$.

    Dispersion measurements are performed in the usual way, by varying the RF
    frequency by a known amount, which changes the beam energy, and
    measuring the change in closed orbit. A standard dispersion measurement at
    CESR varies the 500~MHz  cavities by $\pm 2$~kHz
    (corresponding to $\delta_E/E \pm 5.9\times 10^{-4}$)
    and takes several minutes to acquire. The measurement time is
    determined by the slew rate of the RF frequency of the
    superconducting cavities. The measurement
    reproducibility is better than $5~\textrm{mm}$.

    Quadrupole focusing errors are determined by measuring betatron
    phase advance at each BPM, using turn-by-turn data acquired while
    resonantly exciting the beam \cite{PRSTAB3:092801}. Resonant
    excitation is achieved through
    a pair of ``tune trackers,'' which are stripline kickers
    phase-locked to the horizontal and vertical betatron tunes
    \cite{PAC11:MOP215}. The tune trackers excite the beam to
    amplitudes of several millimeters. Phase and amplitude data are
    extracted from 40,960 consecutive turns by
    a processor onboard the BPM module for each button. The
    button-by-button phase and amplitude are post-processed into
    horizontal and vertical phase, the out-of-phase component
    of the coupling matrix $\bar{C}_{12}$, and the two in-phase
    components of the coupling matrix $\bar{C}_{22,11}$. All of the above
    information is processed from one measurement of the machine.
    Betatron phase data for all 100 BPMs is collected and analyzed in 10
    seconds.
     Reproducibility of betatron phase
    measurements is of order $0.1~\textrm{deg}$.

    When characterizing coupling, only the coupling matrix
    element $\bar{C}_{12}$ is used, and the other two measured
    components ($\bar{C}_{22}$ and $\bar{C}_{11}$) are
    neglected. It is not possible to measure $\bar{C}_{21}$.
    $\bar{C}_{12}$ represents the out-of-phase propagation of coupling
    (a sine-like vertical motion at the horizontal tune, when the horizontal is
    excited with a cosine-like signal), and it can be shown that $\bar{C}_{12}$
    is insensitive to BPM rotations.
    Second, two independent measurements of $\bar{C}_{12}$ can be made
    simultaneously, from excitations of the horizontal and vertical modes,
    adding redundancy in the measurement.
    Third, because the (normalized) coupling matrix elements
    mix from one BPM to the next, measuring and correcting $\bar{C}_{12}$
    globally is sufficient to correct the entire coupling matrix.

    For clarity it is perhaps worth comparing the betatron phase and coupling
    technique with ORM measurement and analysis. The response matrix is
    established by measurement of the closed orbit (position) at each BPM in
    response to excitation of the distributed steering corrector
    magnets. From the set of measured closed orbits (two for each
    steering), the linear transfer matrix (betatron phase advance and
    coupling) from one BPM to the next may be reconstructed.

    In the phase and coupling measurement as implemented for CESR, the
    transfer matrices are similarly reconstructed from measured
    trajectories. The turn by turn capability of the beam position
    monitors is exploited to significantly reduce the measurement
    time. Rather than drive DC correctors to generate a distribution of
    trajectories, the beam is resonantly excited at the betatron
    tunes by a single source at a fixed location.
    As the tunes are non-integer, the 40,960 consecutive trajectories
    smoothly sample phase space, and are acquired in a fraction of a second.
    The trajectories  could in principle be analyzed using ORM
    techniques. Alternatively (and equivalently), we extract betatron
    phase and amplitude and coupling information with the help of
    pre-processing in each of the BPM modules. The objective for the
    emittance correction method at CesrTA is a technique suitable for a
    ring with very large circumference, like the ILC damping ring, therefore
    the betatron phase and coupling technique is favored.

\section{BPM Calibrations}
    In order to ensure that measurements reflect actual machine
    conditions, BPMs must be well-calibrated. The primary
    characteristics to consider are: button-by-button
    timing, button-to-button relative gains, BPM tilts,
    and BPM-to-quadrupole transverse offsets.

    Many modern lightsource BPMs take four signals into one controller
    that pre-processes the raw signals into horizontal and vertical data.
    CESR BPMs have four separate controller cards, one for each button, which
    read out independently. This allows for greater flexibility in measurements
    and post-processing, however some characteristics such as timing and gains
    must be calibrated on each of the four button channels rather than once
    per BPM.

    Each of the required calibrations are now discussed in the order
    of implementation.

    \subsection{BPM Timing}
        Each controller card has independent timing, therefore every
        button on every BPM must be timed in separately.
        A mistimed channel results in sampling the bunch
        passage off-peak, which reduces the observed signal
        amplitude for that button.

        The time-in procedure consists of sampling the temporal profile
        of a bunch passage at a resolution of 10~ps and fitting to determine
        the peak. The process converges in than one minute for all four buttons
        on all 100 BPMs, with less than 10~ps drift over a period
        of four hours.

    \subsection{Button-to-Button Relative Gains}
        Differential response of the four BPM buttons due to variations in
        relative electronic gain will introduce a systematic measurement error.
        Measurements that depend mostly on position, such as orbit, dispersion,
        and the in-phase components of the coupling matrix $\bar{C}_{22,11}$,
        are sensitive to relative button gains. Measurements using relative
        phase, such as betatron phase advance and the out-of-phase coupling
        matrix element $\bar{C}_{12}$ are largely insensitive to gain errors.

        The method of gain mapping used at CesrTA was developed
        by Rubin \emph{et al.} at Cornell \cite{PRSTAB13:092802}, and is based
        on a second-order expansion
        of the button signal response. The method utilizes turn-by-turn data,
        therefore data acquisition is fast, on the order of several
        seconds to collect data for characterizing all 100 BPMs.

        The analysis relies on the fact that a linear relation exists
        between two combinations of the four button signals. For $n$
        turns of turn-by-turn trajectory data there are $4n$ button
        measurements at each BPM. There are only four unknowns, namely
        the button gains, and the system is
        overconstrained for $n > 1$ orbits; typically 1024 turns are used.
        Data acquisition takes about 10 seconds, and the fitting process takes
        less than a minute to determine all four button gains on all 100 BPMs.

        All gain calibration techniques for peak-detection-style BPMs
        are sensitive to timing errors. This method is insensitive to
        detector rotation or offset, as the method uses raw button
        signals across a large cross-section of the BPM, and does not rely
        on distinguishing between horizontal and vertical modes.

        Typical BPM gain variations before correction are of order 5\%, and are
        calibrated with a reproducibility of a few tenths of a percent.

    \subsection{BPM Electronic Centering}\label{sec:centering}

        A relative offset between the electronic center of a BPM and the
        magnetic center of the nearest quadrupole will
        appear in measurements as an offset in the quad. If the relative offset
        is not calibrated, steering the beam to the electronic center of the
        BPM will result in kicks from the quadrupole, generating dispersion.
        To minimize vertical dispersion (and thus the emittance)
        generated during orbit correction,
        the relative offset between the electronic center of a BPM and the
        magnetic center of the nearest quadrupole must be measured.

        The method used at CESR for determining the BPM-to-quadrupole
        transverse offset is based on a common technique where the beam is
        steered such that a change in the quadrupole strength results in no
        change in orbit \cite{PAC95:2452-2454, PAC07:3648-3650}.
        The method employed at CESR has the additional benefit that
        the change in quadrupole strength $K_1$ is determined using betatron
        phase measurements \cite{IPAC10:THPE046}. By measuring the difference
        in phase before and after the quadrupole change, the change in
        quadrupole strength can be more accurately determined than using the
        change in quadrupole current, which is susceptible to hysteresis.
        Therefore, fewer iterations are required on each BPM/quadrupole pair
        to achieve convergence.

        Typical BPM-to-quadrupole offset measurements are around
        $1\textrm{mm}$ RMS in both horizontal and vertical, with a
        short-term reproducibility of order $50~\mu$m and long-term drift of
        about $110~\mu\textrm{m}$ over the course of a three-week
        machine studies period.

        BPM-to-quadrupole relative centering will only affect orbit
        measurements and turn-by-turn trajectory data. Dispersion
        measurements are a difference of two closed orbits,
        therefore absolute offsets do not affect the measurement. Betatron
        phase and coupling measurements are computed button-by-button,
        therefore transverse offsets will not affect the measurement.

    \subsection{BPM Tilt Calibration}

        If a BPM is rotated, a horizontal orbit perturbation will indicate
        a vertical offset. This becomes particularly significant when measuring
        the dispersion, as the average horizontal dispersion in CESR is large,
        on the order of a meter. The lowest vertical dispersion measured
        without
        BPM tilt corrections is around 12~mm RMS. The measurements cannot be
        fit with a model dispersion function generated by corrector magnets or
        magnet misalignments. Furthermore,
        simulations have shown that an RMS of 12~mm of actual vertical
        dispersion corresponds to $20-30~\textrm{pm}$ vertical emittance,
        significantly
        larger than the emittance determined from vertical beam size
        measurements. This implies the measurement is
        dominated by systematic errors, such as uncalibrated BPM tilts.
        If the BPM tilts are uncorrelated with the horizontal dispersion,
        this constrains the distribution of BPM tilts to have
        a maximum RMS of 12~mrad.

        BPM tilts can in principle be determined by linearly fitting
        turn-by-turn
        trajectory data for a well-decoupled beam which is resonantly excited
        in the horizontal mode.
        Residual in-phase coupling will also rotate the beam in x-y
        space, which introduces a lower bound on the ability to resolve
        BPM tilts using this method to around 5~mrad.
        To date, applying the BPM tilts to dispersion
        data does not improve the ability to fit the vertical
        dispersion function. Several alternative methods for
        measuring BPM tilts have been explored, all yielding
        different calibrations, and none improving the ability to
        fit the vertical dispersion. As such, the tilt calibrations have
        not been utilized during any of the corrections described
        in this paper, and remain the most significant known systematic
        in optics corrections.

\section{Beamsize Instrumentation - xBSM}

    The primary method of determining the effectiveness of vertical emittance
    tuning is direct observation of the vertical beam size, from which the
    emittance can be inferred. CESR is instrumented with two x-ray beam size
    monitors (xBSM), one for each species \cite{PAC11:MOP304, IBIC12:WECD01}.

    The xBSMs are one-dimensional 32-diode arrays with $50~\mu \textrm{m}$
    pixel pitch. The instruments are capable of bunch-by-bunch, turn-by-turn
    measurements with a buffer of 250,000 bunch-turns. Dynamic range for the
    instruments span beam currents $0.25-10~\textrm{mA} = 0.4-16\times
    10^{10}/\,\textrm{bunch}$ at the standard CesrTA operating energy of
    2.085GeV.

    When characterizing low-emittance conditions, the beam is typically imaged
    using a horizontal slit formed by two tungsten blades,
    which act as a one-dimensional pinhole.
    Beam size is determined by fitting to the beam profile over 1024 turns on a
    turn-by-turn basis, then averaging. In this way any effect of turn-by-turn
    beam centroid or x-ray optics
    motion is removed from the measured beam size. The fitting
    procedure takes into account the point-response function (prf) of the
    imaging device (in this case, the 1-dimensional pinhole), including the
    effects of the finite opening angle, depth-of-field, energy spectrum,
    diffraction off the tungsten blades, surface roughness of the tungsten
    blades, and detector response. A detailed analysis of the fitting procedure
    is available in \cite{NIMA:XBSM}. In
    practice, the resolution limit when using the pinhole optics is around
    $10-15~\mu\textrm{m}$. The vertical beta function $\beta_b$ at the xBSM
    source point is approximately 40~m, and the xBSM optics provide a
    magnification of approximately 2.2. Therefore, the xBSMs are able to
    resolve the vertical emittance down to $2.5-5.5~\textrm{pm}$.

\section{Low-Emittance Tuning} \label{sec:let_procedure}
    The low-emittance tuning procedure developed at CesrTA takes
    advantage of the fact that all magnets are independently powered,
    and all BPMs are capable of betatron phase and coupling measurements
    through turn-by-turn acquisition. The procedure is as follows:

    \begin{enumerate}
        \item Measure the closed orbit and correct to a reference orbit (which
        aligns the beam with the xBSM beamline) using all 55 horizontal and
        58 vertical steering correctors.
        \item Measure the betatron phase, transverse coupling ($\bar{C}_{12}$),
        and horizontal dispersion. Fit the model lattice to the measurement
        using all 100 quadrupoles and 27 skew quadrupole correctors,
        and load the computed corrections.
        \item Remeasure the closed orbit, transverse coupling, and vertical
        dispersion. Fit the model lattice to all machine data simultaneously
        using all vertical steerings and skew quadrupoles, and load the fitted
        corrector changes.
    \end{enumerate}

    The turnaround time for one full set of corrections is roughly ten minutes.
    It is standard procedure when first recovering conditions
    to save magnet settings after achieving low emittance, run the
    machine through a well-defined hysteresis loop,
    re-load the previously saved conditions, and repeat the
    emittance tuning procedure to apply minor corrections and ensure
    the desired conditions are reproducible.

    Lattice corrections are determined by a $\chi^2$ minimization where a
    machine model is fit to measurements of the lattice functions, with a
    merit function defined as \cite{PRSTAB3:092801}:

    \begin{eqnarray}
        \chi^2 = \sum_{i} {w_i}^{data} \left[d^{measured}(i) -
        d^{model}(i)\right]^2 + \nonumber\\
        \sum_{j} {w_j}^{var} \left[v^{measured}(i) -
        v^{model}(i)\right]^2
    \end{eqnarray}

    \noindent where $d(i)$ is the $i^{th}$ datum (for example,
    the vertical orbit at a BPM), $v(j)$ is the $j^{th}$
    variable (such as a corrector strength), and $w_{i,j}$ are
    user-defined weights. The merit function is minimized by
    adjusting corrector magnets in the model such that the
    model reproduces the measurements. The negative of the machine model
    corrector strengths are then loaded into the machine to
    compensate for optics errors.

    Beam-based characterization of the machine after a
    typical low-emittance correction is shown in Table
    \ref{tab:let_correction_levels}. The discrepancy between the model that
    best fits those measurements and the design demonstrates the effectiveness
    of the correction.

    \begin{table}[htb]
    \caption{Typical levels of correction for optics measurement after
    the full emittance tuning procedure. Measurements were taken at 0.8~mA
    ($1.3\times 10^{10} / \textrm{bunch}$). RMS deviations from the
    design are reported for
    both the machine measurement and a model of the machine which is fit
    to the measurements. Beta beat is computed from fitting phase data.}
    \label{tab:let_correction_levels}
        \begin{tabular}{ccccc}
            \toprule[1pt]
            \addlinespace[2pt]
            \textbf{Measurement} & \textbf{RMS (Data)} & \textbf{RMS (Model)}
            & \textbf{Units} \\
            \midrule[0.5pt]
            \addlinespace[2pt]
            $\delta y$           & 253   & 110      & $\mu$m\\
            $\delta \phi_{a,b}$  & 0.3   & 0.3      & deg   \\
            $\delta \beta/\beta$ & ---   & $0.73\%$ & \%    \\
            $\eta_y$             & 13    & 5        & mm    \\
            $\bar{C}_{12}$       & 0.004 & 0.003    & --  \\
            \bottomrule[1pt]
        \end{tabular}
    % xBSM RD #'s 30666/67, D-line, from eLog #1476
    \end{table}

    The $b$-mode emittance
    is determined from measurements of the beam size and local optics at the
    beam size monitor source point:

    \begin{eqnarray}
        \epsilon_{b} = \frac{\sigma_{y,b}^2}{\gamma_c^2\beta_b}
        \label{eqn:emit_from_beamsize}
    \end{eqnarray}

    \noindent where $\gamma_c$ is a parameter related to the
      coupling matrix, and is approximately unity when coupling is
      small. $\sigma_{y,b}$ is the contribution from the $b$-mode to the
    vertical beam size, which is computed from the total vertical beam
    size $\sigma_y$:

    \begin{eqnarray}
        \sigma_y &=& \sqrt{\sigma_{y,a}^2 + \sigma_{y,b}^2 +
        \sigma_{y,\eta_y}^2}\label{eqn:beamsize_y_coupled}\\
        \sigma_{y,a} &=& \sqrt{\epsilon_a\;\beta_b} \left[\bar{C}_{22}^2 +
        \bar{C}_{12}^2\right]^{1/2} \label{eqn:sigmay_with_coupling}\\
        %\sigma_{y,b} &=& \gamma_c\sqrt{\epsilon_b\;\beta_b}\\
        \sigma_{y,\eta_y} &=&
        \eta_y\frac{\sigma_E}{E}\label{eqn:sigmay_with_eta}
    \end{eqnarray}

    \noindent where $\sigma_{y,a}$, $\sigma_{y,b}$, and $\sigma_{y,\eta_y}$ are
    the contributions to the vertical beam size from the horizontal emittance,
    vertical emittance, and vertical dispersion. $\epsilon_a$ is the
    horizontal-like normal mode.  $\bar{C}_{22}$ and
    $\bar{C}_{12}$ are in-phase and out-of-phase components of the coupling
    matrix, respectively, and are directly measured at BPMs adjacent to the
    source point. Equations
    \ref{eqn:beamsize_y_coupled}-\ref{eqn:sigmay_with_eta}
    are used to determine the component of the vertical beam size due to
    $b$-mode emittance, $\sigma_{y,b}$, which is then used in Eqn.
    \ref{eqn:emit_from_beamsize}.

    $\eta_y$, $\beta_b$, $\bar{C}_{22}$, and
    $\bar{C}_{12}$  are determined by fitting a model of the accelerator to
    measurements. The beam size at the source point $\sigma_{y}$ is calculated
    from the measured image at the xBSM $\sigma_{im}$, accounting for the
    magnification, energy spectrum, and point-source response of the
    optic/detector system.

    Statistical and systematic errors associated with measurements of
    vertical emittance with the xBSM are
    outlined in \cite{PAC11:WEP022}, and include
    contributions from: turn-by-turn beamsize fitting uncertainty;
    turn-by-turn beamsize fluctuation; uncertainty in pinhole size;
    uncertainty in $\beta$ functions; uncertainty in
    longitudinal location of the x-ray source point; and uncertainty
    in dispersion at the source point. The uncertainties propagate as follows:

    %\vspace{20pt}

    \begin{eqnarray}
    \delta \epsilon_b^{sys} &=& \left|\frac{d\epsilon_b}{d\sigma_{im}}\right|
    \delta \sigma_{im}^{sys} + \left|\frac{d
        \epsilon_b}{d\sigma_p}\right| \delta \sigma_p+ \left|\frac{d
        \epsilon_b}{ds}\right| \delta s \label{eqn:demit_sys}\\
    \delta \epsilon_b^{stat}  &=&
    \left(\left|\frac{\partial \epsilon_b}{\partial \beta_b}\right|^2
    \hspace{-5pt} {\left(\delta \beta_b^{stat}\right)}^2
    \hspace{-3pt}+\hspace{-3pt} \left|\frac{\partial \epsilon_b}
    {\partial \eta_y}\right|^2 \hspace{-5pt}
    {\left(\delta \eta_y^{stat}\right)}^2\hspace{-3pt}\right. \nonumber\\
    & & \left.+ \hspace{-3pt} \left|\frac{\partial \epsilon_b}
    {\partial \bar{C}_{22}}\right|^2 \hspace{-5pt}
    {\left(\delta \bar{C}_{22}^{stat}\right)}^2\hspace{-3pt}
    + \hspace{-3pt} \left|\frac{\partial \epsilon_b}
    {\partial \bar{C}_{12}}\right|^2 \hspace{-5pt}
    {\left(\delta
    \bar{C}_{12}^{stat}\right)}^2\hspace{-3pt}\right.\nonumber\\
    & & + \left.\hspace{-3pt} \left|\frac{\partial \epsilon_b}{\partial
    \sigma_{im}}\right|^2\hspace{-5pt}
    {\left(\delta \sigma_{im}^{stat}\right)}^2\right)^{1/2}
    \label{eqn:demit_stat}
    \end{eqnarray}

    \noindent where

    \begin{eqnarray}
    \left|\frac{d \epsilon_b}{ds}\right| &=& \left|\frac{\partial
    \epsilon_b}{\partial \beta_b} \frac{\partial \beta_b}{\partial s} +
    \frac{\partial \epsilon_b}{\partial \bar{C}_{22}} \frac{\partial
    \bar{C}_{22}}{\partial s}+
    \frac{\partial \epsilon_b}{\partial \bar{C}_{12}} \frac{\partial
    \bar{C}_{12}}
    {\partial s}+\right.\nonumber\\
    & &\left.\frac{\partial \epsilon_b}{\partial \eta_y} \frac{\partial \eta_y}
    {\partial s} + \frac{\partial \epsilon_b}{\partial M}
    \frac{\partial M}{\partial s}\right|\label{eqn:demit_ds}
    \end{eqnarray}

    \noindent and $sys$ and $stat$ refer to the systematic and statistical
    uncertainties, respectively. The individual terms $d\epsilon_b / dx_i$
    are computed by varying the terms $x_i$ in the emittance calculation
    by their uncertainties $\pm \delta x_i$.

    Using the above tuning method, and propagating errors according to
    Equations \ref{eqn:demit_stat}--\ref{eqn:demit_ds}, the vertical
    emittances achieved at CesrTA are reported in Table
    \ref{tab:cesrta_emittance}.

        \begin{table}[htb]
        \begin{centering}
        \caption[Emittance corrections achieved at CesrTA]
        {Lowest-achieved vertical emittance at CesrTA in a variety of energies.
        Electron conditions are only reported for the April 2013 CesrTA run at
        2.085~GeV.}
        \label{tab:cesrta_emittance}
        \begin{tabular}{ccccc}
        \toprule[1pt]
        \addlinespace[2pt]
        \textbf{Energy~[GeV]} & \textbf{Species} &
        $\boldsymbol{\epsilon}_{\mathbf{b}}$~\textbf{(pm)} &
        $\boldsymbol{\delta}\boldsymbol{\epsilon}_\mathbf{b}
        ^{\mathbf{sys}}$~\textbf{(pm)} &
        $\boldsymbol{\delta}\boldsymbol{\epsilon}_\mathbf{b}
        ^{\mathbf{stat}}$~\textbf{(pm)} \\
        \midrule[0.5pt]
        \addlinespace[2pt]
        2.085 (03/2013) & $e^+$ & 10.3 & $\left\{\begin{array}{c} +3.2 \\
        -3.4 \end{array}\right.$ & $\left\{\begin{array}{c} +0.2  \\
        -0.2 \end{array}\right.$  \\%& D-40103\\
        2.085 (03/2013) & $e^-$ & 13.3 & $\left\{\begin{array}{c} +3.3 \\
        -3.4 \end{array}\right.$ & $\left\{\begin{array}{c} +0.3  \\
        -0.3 \end{array}\right.$  \\%& C-30818 \\
        %
        %\midrule[0.5pt]
        %
        2.305 (12/2012) & $e^+$ & 10.0 & $\left\{\begin{array}{c} +2.8 \\
        -3.7 \end{array}\right.$ & $\left\{\begin{array}{c} +0.2  \\
        -0.2 \end{array}\right.$  \\%& D-31925\\
        2.553 (03/2013) & $e^+$ & 10.2 & $\left\{\begin{array}{c} +2.9 \\
        -3.4 \end{array}\right.$ & $\left\{\begin{array}{c} +0.2  \\
        -0.2 \end{array}\right.$  \\%& D-45088 \\
        \bottomrule[1pt]
        \end{tabular}
        \end{centering}
        \end{table}

    These results warrant a few comments. First, all four reported
    measurements are within $1\sigma$ of the target $\epsilon_b <
    10$~pm. It is also interesting to note that the minimum-achieved
    emittances are independent of energy. All four measurements are
    within $1\sigma$ of each other, with only the single electron
    measurement standing out.

    The component of the observed vertical beam size due to
    local coupling ($\sigma_{y,a}$) only
    introduces 0.5~$\mu$m in quadrature. The observed beam size is of
    order 20~$\mu$m, and as such, the contribution of local
    coupling is insignificant.

    It is also worth reminding the reader that the minimum measured
    RMS $\eta_y$ of 12~mm corresponds to a vertical emittance of 20-30~pm,
    much larger than what has been measured. It is clear that the dispersion
    measurement is dominated by systematics, and in particular, BPM tilts must
    be better understood.
    However, correcting the vertical orbit and transverse coupling indirectly
    reduces the vertical dispersion. The procedure therefore results in a
    vertical dispersion below the present resolution of the measurement.

    Several alternative LET tuning methods have been explored,
    including Orbit Response Matrix (ORM) analysis \cite{PAC09:WE6PFP104}
    and normal-mode analysis \cite{PRSTAB14:072804}. To date, no method has
    proven to be faster or yield consistently better results than
    the three-stage correction algorithm based on betatron phase and coupling
    measurements discussed here.

\section{LET Simulations}

    To better characterize what factors are limiting emittance corrections,
    software has been developed to evaluate the contributions of misalignments,
    BPM measurement errors, and choice of correction procedure. The program,
    \texttt{ring\_ma2}, uses the \texttt{Bmad}
    accelerator code library \cite{NIMA558:356to359}, and does the following:

     \begin{enumerate}
        \item Assigns random misalignments and BPM errors with user-defined
        amplitudes to the ideal lattice to create a realistic machine model.
        \item Simulates beam based measurements of optics functions
        including the effects of BPM measurement errors.
        \item Computes and applies corrections for each iteration
        based on the simulated measurements.
        \item After each correction iteration, it records the
        effectiveness of the correction in terms of emittances
        and optics functions.
    \end{enumerate}

    The entire procedure is repeated typically 100 times in order to
    generate statistics for analysis. The simulation is approached from a
    statistical perspective for three reasons. First, magnet positions
    continually drift, making it difficult to know the exact set of
    misalignments in the ring on any given day. Second, the precise
    distributions of magnetic centering or BPM measurement
    errors are not known, mandating that their distributions be approached
    from a statistical perspective. Third, by framing the analysis in terms of
    statistical probability of achieving the required emittance, the
    characterization process may be extrapolated to new machines which
    are not yet built using only the knowledge of survey and alignment
    tolerances.

    When discussing the results of statistical analysis the 95\% confidence
    levels (CL) are presented. That is, after applying the full optics
    correction procedure 95\% of simulated lattices, each with a randomly
    chosen distribution of misalignments and measurement errors, achieve a
    vertical emittance below the 95\%CL. The simulation is believed to be
    sufficiently complete such that it is very unlikely that the contribution
    of the static optics to the vertical emittance is greater than this number.

    In this section the method for simulating optics measurements is discussed,
    including how BPM measurement errors and guide field magnet errors are
    modeled. Results of simulations based on input parameters representing the
    physical accelerator are given.

    %-------------------------------------------------------------------
    \subsection{Model Lattice with Errors}
        \texttt{Bmad} allows for introducing strength errors (including
        systematic and random multipole errors) and alignment errors
        (such as offset, roll, and pitch) to any lattice element.
        Magnet strength errors scale
        with the absolute strength of the magnet. Alignment errors are
        treated as additive errors, and are applied directly without scaling.

        \texttt{ring\_ma2} also models BPM measurement errors,
        which are discussed in detail in Section \ref{sec:bpm_sim}.

    %-------------------------------------------------------------------
    \subsection{Simulated Measurements}
        All simulated measurements are modeled as realistically as possible.
        For closed orbit measurements this involves recording 1024 turns of
        trajectory data, including the effects of BPM measurement errors on
        every turn, and averaging the results. Dispersion is simulated as a
        difference of two closed orbits, varying the RF frequency
        in-between.

        For phase and coupling measurements, a particle is resonantly
        excited using a simulated phase-locked tune tracker and allowed to
        equilibrate by tracking for several damping times ($10^5$ turns).
        After the particle trajectory has equilibrated, 40,960 turns of raw BPM
        button data are recorded at every BPM. The data is then processed
        with the same code used for processing CESR phase and coupling data.

        A comparison of lattice parameters derived from simulated measurements
        in an ideal lattice and those computed directly are summarized
        in Table \ref{tab:sim_ideal} for each measurement type, and presumably
        represent a fundamental lower limit to the resolution of each
        measurement technique given no errors in the BPM measurements.
        Simulated measurements have differing levels of agreement for
        horizontal and vertical, which can be attributed to energy loss from
        stochastic radiation emission, leading to a ``sawtooth'' horizontal
        dispersive orbit between the two pairs of RF cavities on opposite
        sides of the ring. This effect is not seen in the vertical as there
        is no vertical dispersion in the design lattice.

        \begin{table}[htb]
        \begin{centering}
        \caption[Simulated measurements for ideal CesrTA lattice]
        {RMS difference between simulated measurements and
        \texttt{Bmad}-calculated values, neglecting any BPM measurement
        errors.}\label{tab:sim_ideal}
        \begin{tabular}{ccc}
            \toprule[1pt]
            \addlinespace[2pt]
            \textbf{Measurement} & \textbf{RMS (Simulated - \texttt{Bmad})} &
            \textbf{Units} \\
            \midrule[0.5pt]
            \addlinespace[2pt]
            Closed Orbit $x,y$ & 0.70,  $< 1\times 10^{-3}$ & $\mu$m\\
            $\eta_{x,y}$ & 0.75, $< 1\times 10^{-6}$ & mm \\
            $\phi_{x,y}$    & 0.1, 0.05 & deg \\
            $\bar{C}_{12}$ & $4.3\times 10^{-4}$ & --\\
            \bottomrule[1pt]
        \end{tabular}
        \end{centering}
        \end{table}

    \subsection{BPM Errors} \label{sec:bpm_sim}
        To generate simulated measurements as realistically as
        possible, BPM measurement errors must be taken into account.
        The two classes of BPM errors modeled in \texttt{ring\_ma2} are BPM
        misalignments (offsets, tilts, and shear) and button-by-button effects
        (button gain, timing, and electronic noise). Each class of errors
        will affect the measurement differently. All
        simulated measurements presented include the effects of
        all listed BPM measurement errors.

        \subsubsection{BPM Misalignments}
            Errors in BPM misalignments (offsets and tilts)
            are applied in the following way:

            \begin{eqnarray}
                \left(\begin{array}{c} x \\ y \end{array}\right)^{m} =
                R(\theta)
                \left(\begin{array}{c} x^{lab} - \delta x \\ y^{lab} - \delta
                y
                \end{array}\right) \label{eqn:geometric_bpm_errors}
            \end{eqnarray}

            \noindent where $(x,y)^{m}$ are the coordinates with BPM
            misalignments applied, $R(\theta)$ is the rotation matrix for angle
            $\theta$, and $\delta x, \delta y$ are the horizontal and vertical
            offset between the BPM and nearest quadrupole.

        \subsubsection{Button Effects: Gain, Timing, and Reproducibility}
            Timing errors, gain variations, and turn-by-turn resolution
            affect individual button signals. Modeling their effects requires
            an accurate method for converting from $(x,y)$ coordinates to
            button signals $b_{1,2,3,4}$, applying errors, and
            converting back to $(x,y)$ coordinates.

            All button-by-button
            errors of these classes are handled through use of a nonlinear
            interpolation grid which converts $(x,y)$ coordinates to
            button signals. Button-by-button errors are applied to the
            individual channels, and the final ``measured'' $(x,y)$
            coordinates are determined by the best fit to the set of
            new button signals using the same interpolation grid
            \cite{PRSTAB8:062802}. The nonlinear map used in these studies
            is for a BPM with a ``CESR geometry'' (see Fig.
            \ref{fig:cesr_bpm}).

            Including effects from button-to-button gain errors, timing errors,
            and measurement reproducibility, the four observed button signals
            $b_i$ at each BPM are:

            \begin{eqnarray}
                b_i^{meas} = g_i\,t_i\,b_i^{m} + \delta b_i^{noise}
                \label{eqn:bi_meas}
            \end{eqnarray}

            In Equation \ref{eqn:bi_meas} $b_i^{m}$ are defined to be the
            button signals determined through the interpolation grid
            for the coordinates $(x,y)^{m}$ from Equation
            \ref{eqn:geometric_bpm_errors}. $g_i$ is the gain error on button
            $i$, and $t_i$ is an effective gain error for button $i$ arising
            from the timing error:

            \begin{eqnarray}
                t_i = 1 - \frac{a_0}{a_2 + \frac{a_1^2}{4a_0}} \left(
                \delta t[s]\right)^2
            \end{eqnarray}

            \noindent where the constants $a_{0,1,2}$ are empirically
            determined. Note that because CESR BPMs are timed to the
            peak signal of a bunch passage, any timing error will decrease
            the button signal. This method of modeling the
            timing error also allows the BPM model to account for
            synchrotron motion, thus modulating the timings on
            all four buttons on a turn-by-turn basis.

            BPM position measurement reproducibility is dominated by electronic
            noise arising from the digitization and amplification of an analog
            signal on each of the four controllers, and is modeled in
            Equation \ref{eqn:bi_meas} as an additive error
            $\delta b_i^{noise}$ on each of the four button signals.
            The amplitude of the button-by-button reproducibility is
            set by determining the change in a single button signal consistent
            with changing the observed orbit by the desired reproducibility
            (for example $7\mu$m).

            The top and bottom CESR BPM button blocks are welded separately
            to the vacuum chamber. There is then the possibility of a relative
            misalignment of the two blocks. In order to estimate the effect of
            this misalignment in simulation, upper and lower button signals
            are determined by offsetting the BPM in opposite directions.

    %-------------------------------------------------------------------
    \subsection{Simulation Results}
        Amplitudes for misalignments and BPM errors in the simulation are
        summarized in Tables \ref{tab:cesrta_sim_bpm_errors},
        \ref{tab:cesrta_sim_misalignments}, and
        \ref{tab:cesrta_sim_multipoles}, and are determined either from
        directly-measured values or inferred from machine measurements.
        Offsets of quadrupoles and sextupoles
        include measured alignment levels along with 100~$\mu$m added in
        quadrature to account for the estimated uncertainty in the offset
        of magnetic center with respect to geometric center of these elements.

\begin{figure}[tbh]
        \includegraphics[width=3 in]{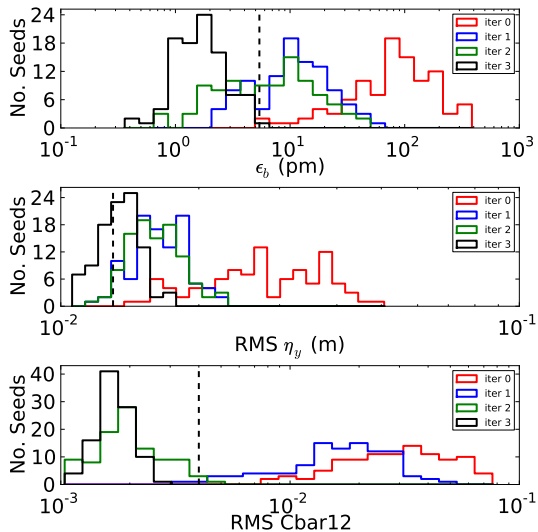}
        \caption{Results from \texttt{ring\_ma2}, using
        misalignments, BPM measurement errors, and multipoles stated in
        Tables \ref{tab:cesrta_sim_bpm_errors},
        \ref{tab:cesrta_sim_misalignments}, and
        \ref{tab:cesrta_sim_multipoles},
        plotted before correction (red), and after first, second, and third
        stage of emittance correction (blue, green, and black, respectively).
        The dashed black line indicates typical measured values in CESR after
        low-emittance correction.}
        \label{fig:ringma2_results}
        \end{figure}

        The emittance correction procedure used in the simulation is
        identical to that used on the actual machine, outlined in
        Section \ref{sec:let_procedure}. Results from
        \texttt{ring\_ma2} are shown in Fig.
        \ref{fig:ringma2_results}, and summarized in Table
        \ref{tab:let_sim_results}.

        % /home/shanksj/CesrTA/ring_ma2_jobs/
        % 20130701_thesis_omni/baseline/tracking/*
        \begin{table}[htb]
        \caption{95\% confidence level (CL) correction levels after each
        correction iteration. All values except ${\eta_y}^{Bmad}$ include
        observational effects from BPM measurement errors. Details of the
        correction iterations are discussed in Section
        \ref{sec:let_procedure}.}\label{tab:let_sim_results}
        \begin{tabular}{cccccc}
            \toprule[1pt]
            \addlinespace[2pt]
            \textbf{Measurement}     & \textbf{Initial} & \textbf{Iter 1} &
            \textbf{Iter 2} & \textbf{Iter 3} & \textbf{Units}   \\
            \midrule[0.5pt]
            \addlinespace[2pt]
            $\phi$            &  7.7    & 1.6  & 0.1  & 0.1  & deg\\
            ${\eta_y}^{Meas}$ &  42.6   & 18.7 & 18.7 & 15.4 & mm\\
            ${\eta_y}^{Bmad}$ &  40.1   & 13.9 & 12.2 & 5.0  & mm\\
            $\bar{C}_{12}$    &  6.3    & 3.2  & 0.34 & 0.24 &
            $\times 10^{-2}$ \\
            $\epsilon_b$      &  255.8  & 33.0 & 27.5 & 4.1  & pm\\
            \bottomrule[1pt]
        \end{tabular}
        \end{table}

        After correction, 95\% of seeds achieved a vertical
        emittance below 4.1~pm, which is significantly smaller than the minimum
        measured vertical emittance of 10.3 (+3.2/-3.4)$^{\;sys}$
        ($\pm$0.2)$^{stat}$~pm at 2.085~GeV. It is clear that the simulation
        does not account for more than half of the measured vertical
        emittance.

        BPM tilts are the single most significant contribution to the
        vertical emittance in the simulation, and dominate the simulated
        $\eta_y$ measurement. Considering the simulation under-estimates the
        measured vertical emittance, one could envision adjusting the
        simulation
        to reduce the amplitude of BPM tilts and increase magnet misalignments
        in order to increase the vertical emittance to levels measured in
        the actual machine while holding the RMS $\eta_y$ constant.
        However, the required change in alignment to generate the measured
        vertical emittance is much larger than the measured uncertainty in
        the alignment procedure.

\section{Diagnosis of Emittance Dilution}

        The two primary mechanisms for the
        static optics to contribute to vertical emittance are vertical
        dispersion and horizontal-to-vertical coupling. The measured vertical
        dispersion in Table \ref{tab:let_correction_levels} and the minimum
        $\bar{C}_{12}$ measured at CesrTA ($2\times 10^{-3}$) are within the
        distributions from the simulation, indicating that the model is
        realistic. Increasing the coupling in simulated lattices such that the
        $\bar{C}_{12}$ RMS is consistent with the measurement in Table
        \ref{tab:let_correction_levels} introduces less than 1~pm of vertical
        emittance.

        Additionally, significant efforts have been made to ensure that
        all sources of uncertainty in the emittance measurement are
        accounted for. The discrepancy is therefore not attributed to emittance
        measurement errors. This implies there are significant
        sources of vertical emittance that are not included in the model or
        \texttt{ring\_ma2} simulation.  Potential sources of vertical
        emittance are now considered.

    \subsection{Emittance Dilution from RF}

        Random RF voltage and phase jitter may contribute to
        emittance dilution.
        There are four superconducting RF cavities in CESR, split into
        two pairs. Each pair is powered by a single power supply.
        Turn-by-turn beam size was recorded while varying
        the total RF voltage and number of RF cavities powered.
        The results are summarized in Table~\ref{tab:rf_tests}.
        It should be noted that the studies summarized in this section were
        taken while one of the two West RF cavities was disabled, therefore
        only three RF cavities were used (one in the West, and two in the
        East). Nominal total RF voltage was 4.8~MV, distributed approximately
        evenly among the three cavities.

        \begin{table}[htb]
            \begin{centering}
            \caption[Summary of beam stability tests at CesrTA]
            {Summary of beam stability tests at CesrTA. The measurements
            were conducted in April 2013, for a single bunch of positrons
            at 0.7-0.85~mA.}
            \label{tab:rf_tests}
            \begin{tabular}{cccc}
            \toprule[1pt]
            \addlinespace[2pt]
            \textbf{Total RF (MV)} & \textbf{East RF} & \textbf{West RF}  &
            $\boldsymbol{\epsilon}_\mathbf{b}$~\textbf{(pm)} \\
            \midrule[0.5pt]
            \addlinespace[2pt]
            4.8 & On  & On  & 11.5  \\
            1.7 & On  & On  & 11.2  \\
            1.7 & Off & On  & 12.5  \\
            1.7 & On  & Off & 10.8  \\
            \bottomrule[1pt]
            \end{tabular}
            \end{centering}
        \end{table}

        A small reduction in beam size was observed when
        reducing the total RF voltage from 4.8~MV to 1.7~MV, corresponding to
        a reduction in observed vertical emittance of 0.3~pm. The $1\sigma$
        statistical uncertainty for the lowest-measured emittance is
        $\pm$0.2~pm. Note that although the systematic uncertainty is an
        order of magnitude larger, it represents a global uncertainty where
        all measurements would be affected uniformly by any change in the
        understanding of the beamsize measurement system.

        A further reduction is seen when the single West RF cavity is powered
        down and detuned, such that only the two East RF cavities are running;
        the emittance increased slightly when running only on the W1 RF cavity.
        This indicates that the RF system is contributing to the vertical
        emittance, although the mechanism is not known at this
        time.
        The East and West RF cavity pairs run on separate power supplies;
        one hypothesis is that the West RF power supply is less stable than the
        East, thereby introducing vertical emittance through modulation of the
        RF voltage. Alternatively, by running a single cavity at a higher
        voltage, the amplitude of voltage jitter is also increased, potentially
        increasing the contribution to the emittance. The RF system in CESR is
        superconducting, therefore a direct examination of the alignment
        requires the nontrivial process of warming the cavities and opening the
        cryostats.

    \subsection{Collective Effects}

        The CesrTA emittance target of 10~pm is for a ``zero-current'' beam;
        that is, neglecting any collective effects. Effects
        considered here include: electron cloud, fast-ion
        instability, intra-beam scattering, and wakefields.

        Electron cloud and fast-ion instability typically require
        a train of 30 bunches with 0.5~mA/bunch or more in order for
        emittance dilution to occur, and the
        emittance blow-up takes place around bunch 10-15 in the train
        \cite{IPAC13:TUPWA063}. A single bunch is not capable of generating
        sufficient cloud or ion density to cause emittance dilution.

        Extensive measurements and simulations of intra-beam scattering (IBS)
        at CesrTA indicate that the vertical emittance is largely insensitive
        to IBS effects at currents $I < 1$~mA/bunch, where the measurements
        reported here were taken \cite{CornellU2013:PHD:MPEhrlichman,
        IPAC12:WEPPR015}. The mechanism through which IBS increases vertical
        emittance depends on either transverse-to-longitudinal scattering in
        regions with dispersion or transverse-to-transverse scattering in
        regions with coupling, such that the vertical-mode action of the
        particle changes. Vertical dispersion and coupling are measured to be
        globally well-corrected, and are well below levels required for IBS to
        contribute to vertical emittance dilution.

        Wakefields would tend to increase the emittance
        linearly with current.
        By measuring the dependence of the vertical beam size
        on current, it may be
        possible to determine whether wakefields are contributing
        to the emittance at the nominal 0.8~mA/bunch used for the emittance
        measurements presented in Table \ref{tab:cesrta_emittance}.
        However, at such low current, photons are sparse and the
        turn-by-turn fitting procedure is no longer reliable.
        Instead, the turn-by-turn images must be
        averaged first in order to improve signal-to-noise, then fit as a
        single image. This has the disadvantages of incorporating a small
        amount of turn-by-turn beam motion and increasing the statistical
        uncertainty in the vertical emittance measurement.

\begin{figure}[tbh]
        \begin{centering}
            \includegraphics[width=3.5 in]{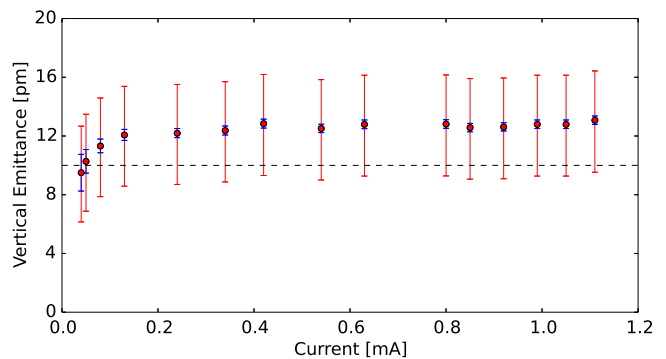}
            \caption[Vertical emittance as a function of bunch current]
            {Vertical emittance for a single bunch of positrons
            as a function of bunch current, from April 2013
            CesrTA machine studies. Plotted error bars are systematic (red)
            and statistical (blue). The dashed horizontal line indicates the
            10~pm zero-current vertical emittance target for CesrTA.}
            \label{fig:emit_vs_current}
        \end{centering}
        \end{figure}

        Figure \ref{fig:emit_vs_current} shows the emittance calculated from
        a series of vertical beam size measurements from the xBSM, taken
        sequentially as the current was decreased from 1.1~mA to around
        0.05~mA, and processed as described above.

        The rolloff of beam size at very low current ($< 0.1$~mA) is
        likely due to lack of sufficient flux on the beam size
        monitor. Moreover, the emittance does not depend linearly on beam
        current. As such, there is no support for current-dependent
        effects diluting the vertical emittance at low-current.

\section{Conclusions}

    A low-emittance tuning procedure has been developed at CesrTA, based on
    betatron phase and coupling measurements using resonant excitation and
    turn-by-turn capable BPM. The tuning procedure has a fast turnaround,
    where one round of optics correction takes about ten minutes, and has
    yielded a single-bunch vertical emittance of $\epsilon_y = 10.3$
    (+3.2/-3.4)$^{\;sys}$ ($\pm$0.2)$^{stat}$~pm with a single bunch of
    positrons with $0.8~\textrm{mA} =
    1.3\times10^{10}$ at 2.085~GeV. The correction procedure routinely
    achieves $\epsilon_y < 15$~pm in a variety of machine conditions at
    energies ranging from 2.085-2.5~GeV.

    The tuning procedure developed at CesrTA is significantly
    faster than response matrix analysis. The method scales
    independently of number of BPMs or correctors, thus for large
    machines the CesrTA procedure will be proportionally faster than
    response matrix analysis.

    The lack of energy dependence for the minimum-achieved
    vertical emittance may yield information regarding sources of
    emittance dilution. Further studies, including measuring the
    emittance at several energies during a single machine studies
    period, will be necessary before a conclusive statement may be
    made.

    Collective effects do not appear to contribute to
    emittance dilution for a single bunch at 0.8~mA. The RF
    system on the other hand clearly does affect the emittance,
    and further investigations are necessary.

    Although misalignments do not appear to be the most significant
    contribution to the emittance, any improvement in alignment or
    optics correction will likely result in a small reduction in the emittance,
    as contributions to the emittance add linearly. In particular, BPM tilts remain
    a significant outstanding issue which limits the understanding of $\eta_y$.
    Simulations suggest that a reduction in RMS BPM tilt
    from 12~mrad to 5~mrad could reduce the contribution of the static
    optics to the vertical emittance by 50\%.
    Alternative BPM tilt fitting techniques are under development.

\begin{acknowledgments}
    The authors wish to thank the CESR operations and instrumentation groups,
    whose support was indispensable in our efforts to achieve well-corrected
    conditions. This work was supported by the National Science Foundation
    grant PHY-1002467 and the Department of Energy grant DE-SC0006505.
\end{acknowledgments}

\appendix

\setcounter{secnumdepth}{0}

\section{Appendix: Errors for \texttt{ring\_ma2} Simulations}
\label{app:ringma2_errors}

    Table \ref{tab:cesrta_sim_misalignments} shows the misalignments and
    errors used in CesrTA \texttt{ring\_ma2} studies.
    Offsets of quadrupoles and sextupoles
    include measured alignment levels along with 100~$\mu$m added in
    quadrature to account for the estimated uncertainty in the offset
    of magnetic center with respect to geometric center of these elements.

    Systematic multipoles are included for sextupoles which have vertical
    steering or skew quadrupole trim windings. These multipoles are
    computed using field modeling software, and are scaled to a measurement
    radius of 20~mm. There is a known random skew quadrupole component
    to the damping wiggler fields \cite{PAC05:Crittenden}, due to manufacturing
    tolerances in the radii of the pole windings, which is also included.
    Multipoles used in this study are summarized in Table
    \ref{tab:cesrta_sim_multipoles}, and use the following convention
    (summarized in the \texttt{Bmad} manual \cite{bmad:manual}):

    \begin{eqnarray}
        \frac{qL}{P_0}\left(B_y + \imath B_x\right) =
        \sum_{n=0}^{\infty}\left(b_n + \imath
        a_n\right) \left(x + \imath y\right)^n
    \end{eqnarray}

    \noindent where $b_n$ and $a_n$ are the normal and skew multipoles,
    respectively. The values in the table are normalized by $1/(K_mL r_0^m)$,
    where $m$ is the order of the primary field ($m = 1$ for quadrupole, $2$
    for sextupole, etc.).

    \begin{table}[htb]
    \begin{centering}
        \caption [CesrTA BPM errors for \texttt{ring\_ma2}]
        {BPM errors introduced into model
        CesrTA lattice for \texttt{ring\_ma2} studies.}
        \label{tab:cesrta_sim_bpm_errors}
        \begin{tabular}{ccc}
            \toprule[1pt]
            \addlinespace[2pt]
            \textbf{Error} & \textbf{Applied RMS} & \textbf{Units}\\
            \midrule[0.5pt]
            \addlinespace[2pt]
            Reproducibility & 10   & $\mu$m  \\
            Tilt            & 12   & mrad    \\
            Gains           & 0.5\%& \%      \\
            Timing          & 10   & ps      \\
            Offset ($x,y$)    & 170  & $\mu$m  \\
            Horizontal Shear  & $\pm 100$  & $\mu$m  \\
            \bottomrule[1pt]
        \end{tabular}
    \end{centering}
    \end{table}

    \begin{table}[htb]
    \begin{centering}
        \caption [CesrTA misalignments and errors for \texttt{ring\_ma2}]
        {Misalignments and errors introduced into model
        CesrTA lattice for \texttt{ring\_ma2} studies.
        All parameters are determined either from machine
        measurements or survey.}
        \label{tab:cesrta_sim_misalignments}
        \begin{tabular}{rrrc}
            \toprule[1pt]
            \addlinespace[2pt]
            \textbf{Element Class} & \textbf{Error} & \textbf{RMS} &
            \textbf{Units} \\
            \midrule[0.5pt]
            \addlinespace[2pt]
            Dipole      & $x$ Offset & 0.9  & mm       \\
                        & $y$ Offset & 2.0  & mm       \\
                        & $s$ Offset & 2.3  & mm       \\
                        & Roll       & 144  & $\mu$rad \\
                        & $x$ Pitch  & 600  & $\mu$rad \\
                        & $y$ Pitch  & 300  & $\mu$rad \\
            \midrule[0.5pt]
            Quadrupole  & $x$ Offset & 335  & $\mu$m   \\
                        & $y$ Offset & 40.3 & $\mu$m   \\
                        & Magnetic Offset & 100 & $\mu$m \\
                        & $s$ Offset & 5.2  & mm       \\
                        & Tilt       & 148  & $\mu$rad \\
                        & $x$ Pitch  & 1100 & $\mu$rad \\
                        & $y$ Pitch  & 62   & $\mu$rad \\
                        & k1         & 0.1\%& \%       \\
            \midrule[0.5pt]
            Sextupole   & $x$ Offset & 280  & $\mu$m   \\
                        & $y$ Offset & 280  & $\mu$m   \\
                        & Magnetic Offset & 100 & $\mu$m \\
                        & $s$ Offset & 5.2  & mm       \\
                        & Tilt       & 200  & $\mu$rad \\
                        & $x$ Pitch  & 1200 & $\mu$rad \\
                        & $y$ Pitch  & 800  & $\mu$rad \\
                        & k2         & 0.1\%& \%       \\
            \midrule[0.5pt]
            Wiggler     & $x$ Offset &   1  & mm       \\
                        & $y$ Offset & 250  & $\mu$m   \\
                        & $s$ Offset & 500  & $\mu$m   \\
                        & Tilt       & 300  & $\mu$rad \\
                        & $x$ Pitch  & 200  & $\mu$rad \\
                        & $y$ Pitch  & 250  & $\mu$rad \\
            \bottomrule[1pt]
        \end{tabular}
    \end{centering}
    \end{table}

    \begin{table}[htb]
    \begin{centering}
        \caption [Multipoles used in \texttt{ring\_ma2} studies of CesrTA
        lattice]
        {Multipoles used in \texttt{ring\_ma2} studies of CesrTA
        lattice. Sextupole multipoles are systematic and therefore identical
        at all sextupoles, whereas the wiggler $a1$ multipole is random;
        the number quoted for wiggler $a1$ is therefore the RMS of the
        applied distribution.}\label{tab:cesrta_sim_multipoles}
        \begin{tabular}{rcr}
            \toprule[1pt]
            \addlinespace[2pt]
            \textbf{Element Class} &
            \textbf{Multipole} &
            \textbf{Value} \\
            \midrule[0.5pt]
            \addlinespace[2pt]
            Sextupole with      & a3  & $-7.25\times 10^{-4}$ \\
            Vert. Steering Trim & a5  & $-1.46\times 10^{-2}$ \\
                                & a7  & $ 6.68\times 10^{-4}$ \\
                                & a9  & $8.7\times 10^{-6}$   \\
                                & a11 & $1.0\times 10^{-5}$   \\
            \midrule[0.5pt]
            Sextupole with      & a4  & $-1.2145\times 10^{-1}$ \\
            Skew Quad Trim      & a6  & $2.16\times 10^{-4}$  \\
                                & a8  & $4.96\times 10^{-4}$  \\
                                & a10 & $-2.29\times 10^{-5}$ \\
                                & a12 & $-1.0\times 10^{-5}$  \\
            \midrule[0.5pt]
            Wiggler                         & a1  & $2.88\times 10^{-4}$ \\
            \bottomrule[1pt]
        \end{tabular}
    \end{centering}
    \end{table}

\cleardoublepage
\bibliography{CesrTA}

\end{document}